\documentclass[epj,referee]{svjour}
\usepackage{graphics}
\begin{document}
\title{Phase-ordering dynamics of binary mixtures with field-dependent
mobility in shear flow}
\author{G. Gonnella, A. Lamura and  D. Suppa
}                     
%
%
\institute{{\it Istituto Nazionale per la Fisica della Materia (INFM),
 Unit\`a di Bari,} and\\
{\it Dipartimento di Fisica dell'Universit\`{a} di Bari,} and
{\it Istituto Nazionale di Fisica Nucleare, Sezione di Bari\\
via Amendola 173, 70126 Bari, Italy}}
\date{Received: date / Revised version: date}
%
\abstract{
The effect of shear flow on the phase-ordering dynamics of
a binary mixture with field-dependent mobility is investigated. The problem
is addressed in the context of the time-dependent Ginzburg-Landau equation
with an external velocity term, studied in self-consistent approximation.
Assuming a scaling ansatz for the structure factor,
the asymptotic behavior of the observables in the
scaling regime can be analytically calculated.
All the observables show log-time periodic oscillations which we 
interpret as due to a cyclical mechanism of stretching and break-up of domains.
These oscillations are dumped as consequence of the vanishing 
of the mobility in the bulk phase.
\PACS{
      {47.20Hw}{Morphological instability; phase changes}   \and
      {05.70Ln}{Nonequlibrium and irreversible thermodynamics}   \and
      {83.50Ax}{Steady shear flows}
     } 
} 
\authorrunning{G. Gonnella {\it et al.}}
\titlerunning{Phase-ordering dynamics of binary mixtures with field-dependent
mobility in shear flow}
\maketitle
\section{Introduction}
A well-studied problem in non-equilibrium statistical mechanics is the
growth of domains in quenching processes \cite{Gunton}. Namely, when a system
is suddenly quenched from a disordered initial state into a thermodynamic 
region where different phases coexist, macroscopic domains can be observed,
usually characterized by a 
single time-dependent length 
scale, which grows as a power law
$L(t) \sim t^{z}$. The spatial patterns of the domains
at two different times are related by a global change of this length scale. A
signature of this dynamical scaling is the fact that the structure factor
 $C(\vec{k},t)$ can 
be cast in the form
\begin{equation}
C(\vec{k},t) = L^{d}(t) C_{o} (\vec{k} L(t))
\label{prima}
\end{equation}
where $C_{o}$ is the scaling function.

In the case of a binary mixture, the evolution of the system is described by
a scalar order parameter $\phi$ representing the difference of concentrations
of the two liquids in the mixture. When hydrodynamic effects are not 
considered, the quenching process can be described by an equation
of the form
$\partial_{t} \phi= \!- \vec{\nabla}\cdot
\vec{j}$, where $\vec{j}=\! -\Gamma \vec{\nabla}(\delta H/\delta \phi)$,
$H$ is a free-energy functional describing the ordered phases and
$\Gamma$ is a transport coefficient called mobility, usually taken constant.
It has been argued \cite{Langer} that 
the dynamics is more accurately mimicked considering a 
field-dependent mobility of
the form  
\begin{equation}
\Gamma(\phi) = (1 - a(T) \phi^{2})
\label{mobility}
\end{equation}
where $a(T) \rightarrow 1$ for temperature $T \rightarrow 0$, 
$a(T) \rightarrow 0$ for $T \rightarrow T_{c}$.
 In this way it is possible to take into account the 
different nature of the two mechanisms which operate during the phase 
separation: surface diffusion and bulk diffusion. The first has a
growth exponent $z=1/4$ \cite{Furukawa} and is due to the diffusion of
 molecules of the two
species along the interfaces. The second,
which is also called Lifshitz-Slyozov mechanism, is due to the diffusion of molecules of
one species from more curved interfaces, where they evaporate, 
to less curved ones through the bulk
of the other phase. The corresponding
growth exponent is $z=1/3$ \cite{Lifshitz}. At high
temperatures (but still less than the critical temperature) the bulk 
diffusion is the only observed because it is faster than the other. 
Since bulk diffusion is a thermal activated process \cite{Corberi}, one
expects that, 
lowering
the temperature, a regime with $z=1/4$ can be observed.
 The proposed form
(\ref{mobility}) for the mobility is able to catch these features. For 
shallow quenches ($a \ll 1$) the mobility remains constant in the whole system,
while for very 
deep quenches ($a=1$) $\Gamma (\phi)$ vanishes in the bulk phases 
where $\phi^{2}=1$, suppressing the bulk diffusion. The diffusion along the
interfaces is unaffected because $\Gamma (\phi) \simeq 1$ on domain boundaries.

The effect of an order parameter-dependent mobility (\ref{mobility}) on  
systems with scalar order parameter has been studied 
by simulations \cite{Lacasta,Puri}.
It has been found that for $a=1$ the length scale $L(t)$ grows as $t^{1/4}$ and
for $0<a<1$ there is a crossover between $L(t) \sim t^{1/4}$ and 
$L(t) \sim t^{1/3}$. Recently, a non constant mobility has been also 
used to study
the phase-ordering dynamics in systems with a vectorial order parameter
\cite{Corberi,Bray}. In \cite{Corberi} the expression (\ref{mobility})
for the mobility has been adopted. The limit $N \rightarrow \infty$ 
, where $N$ is the number of vectorial components, is 
analytically solvable and corresponds to the self-consistent approximation
of the scalar case. It is characterized by a growth exponent $1/6$
for the case $a=1$. For $0<a<1$ the usual
value $1/4$ of the growth exponent for vectorial systems is recovered in the
asymptotic regime. A more general form of the mobility has been 
introduced in \cite{Bray}, being given by
\begin{equation}
\Gamma(\phi) = (1 - \phi^{2})^{\alpha}
\label{newmob}
\end{equation}
where $\alpha$ is a positive real number.

In this paper we are concerned with the phase separation of a binary mixture 
subject to an uniform shear flow when the mobility is given by Eq.~ 
(\ref{newmob}). When a shear flow is applied to a quenched binary mixture,
the pattern of the phase-separating domains as well as the time evolution
are strongly modified by the flow \cite{Onuki}. 
Domains greatly elongated in the flow direction 
have been observed in simulations 
\cite{Ohta,Rothman} and in experiments \cite{Hashimoto}. A difference
$\Delta z = 0.8 \div 1$ between
growth exponents 
has been measured in some experiments,
with the larger exponent in the flow
direction \cite{Lauger,Chan}.
The deformations of domains affects the rheological properties of the system,
giving rise to excess stresses and to
 an increase $\Delta \eta$ of the viscosity \cite{Onuki2,Krall}. This
behavior has its explanation in the fact that mechanical energy is expended to
deform the domains against the interfacial tension.
When the domains are stretched to such an extent that they start to burst,
the stored mechanical energy is dissipated and the excess viscosity
decreases. The phase separation in steady shear for binary mixtures with
constant mobility has been recently studied in \cite{Noi}, where the 
existence of an anisotropic dynamical 
scaling theory with different growth exponents in flow and in
other directions has been shown.
It is found that the excess viscosity, after reaching a maximum, relaxes
to zero, exhibiting log-time periodic oscillations. Also other physical
observables are modulated by such oscillations, which
can be related to a cyclical mechanism of storing and dissipation 
of elastic energy. Here we want to see the effects of a 
non constant mobility on this {\it scenario}.

The outline of the paper is as follows.
In section 2 we present the model and derive the equation of
time evolution for the structure factor in a self-consistent approximation.
In section 3 we report the asymptotic behavior of the model. We found two
different growth exponents for the flow and the shear  directions, given
in the leading scaling regime by $z_{x}=(5+2 \alpha)/2(2+\alpha)$ and
$z_{y}=1/2(2+\alpha)$, respectively. 
The asymptotic behavior of the rheological quantities is also 
calculated. Finally, in section 4 we integrate numerically the time evolution
equation of the structure factor and calculate the whole evolution of the
physical
observables as moments of the structure factor.
In the asymptotic regime they are modulated by dumped
log-time oscillations. Our conclusions complete the article.
\section{The model}
We consider a binary mixture whose evolution is described by the 
diffusion-convection equation
\begin{equation}
{\label{eqM01}}
\frac{\partial \phi(\vec{r},t)}{\partial t}+
\vec{\nabla} \cdot (\phi(\vec{r},t) \vec{v})=
\vec{\nabla} \cdot \left [
\Gamma (\phi) \vec{\nabla} \left ( \frac{\partial
H[\phi(\vec{r},t)]}{\partial \phi}\right ) \right]
\end{equation}
where the field $\phi(\vec{r},t)$ 
describes the concentration difference between the two components
of the mixture, $\Gamma (\phi)$ is
the mobility, which depends on the order parameter as in
Eq.~(\ref{newmob}). The order parameter $\phi$ is convected by an external
velocity field \cite{Onuki}. We choose a planar shear flow with
\begin{equation}
\vec{v}=\gamma y \vec{e}_x,
\end{equation}
where $\gamma$ is the shear rate, assumed constant, and $\vec{e}_x$ is a 
unit vector in the flow direction.
The free energy functional is chosen to be of the standard $\phi^4$
form:
\begin{equation}
{\label{eqM02}}
H[ \phi]= \!\!\int \!\!d \vec{r} \left [ -\frac{1}{2} \phi^2(\vec{r},t)+
\frac{1}{4} \phi^4(\vec{r},t)+\frac{1}{2} |\vec{\nabla}
\phi(\vec{r},t) |^2 \right]
\end{equation}
where we assume that all parameters have been rescaled into dimensionless
units \cite{Sagui} and the system is below the critical temperature.
The two minima of the polynomial part of $H[ \phi]$ describe the
pure states of the mixture.

In this article we deal with the non-linear term of Eq.~ (\ref{eqM01})
in a self-consistent approximation; hence the term $\phi^3$ appearing in
the functional derivative $\delta H/\delta \phi$ is linearized
as $< \phi^2 > \phi$, where $<... >$ stands for the average
over the system. In the same way the term  $\phi^2$ in the mobility is
substituted by  $< \phi^2 >$.
In the Fourier space Eq.~ (\ref{eqM01}) becomes:
\begin{eqnarray}
{\label{eqM03}}
\frac{\partial \hat{\phi}(\vec{k},t)}{\partial t}&=&
\gamma k_x \frac{\partial}{\partial k_y} \hat{\phi}(\vec{k},t) \nonumber \\
&-&
(1-S(t))^\alpha k^2 
\left[ S(t)-1+k^2 \right] \hat{\phi}(\vec{k},t)
\end{eqnarray}
where $S(t)=< \phi^{2}(\vec{r},t) >$ and
$\hat{\phi}(\vec{k},t)$ is the Fourier transform of
$\phi(\vec{r},t)$.

The statistical quantity of experimental interest is the time-dependent
structure factor $C(\vec{k},t)$ which is defined as $<\hat{\phi}(\vec{k},t)
\hat{\phi}(-\vec{k},t)>$. It obeys to the evolution equation:
\begin{eqnarray}
{\label{eqM04}}
\frac{\partial C(\vec{k},t)}{\partial t}&=&
\gamma k_x \frac{\partial}{\partial k_y}
C(\vec{k},t) \nonumber \\
&-& 2 (1-S(t))^\alpha k^2 \left [
S(t)-1+ k^2 \right ] C(\vec{k},t)
\end{eqnarray}
which is closed by the self-consistency condition 
\begin{equation}
S(t)=\int_{|\vec{k}|<q} \frac {d \vec{k}}{(2 \pi)^{d}} \;\: C(\vec{k},t) 
\end{equation}
where $q$ is a phenomenological cutoff.
Rheological quantities of interest can be calculated as momentum integrals
 of the 
structure factor.
Since we  restrict the solution of the model to the two-dimensional case,
we consider
the excess viscosity $\Delta \eta$ and the first normal stress
$\Delta N_{1}$ defined by \cite{Dawson} 
\begin{equation}
\Delta \eta = - \gamma^{-1} \int_{|\vec{k}|<q} \frac{d \vec{k}}{(2 \pi)^{d}} 
\:k_{x}\: k_{y}\: C(\vec{k},t)
\label{eqn5b}
\end{equation}
\begin{equation}
\Delta N_{1} = \int_{|\vec{k}|<q}
 \frac{d \vec{k}}{(2 \pi)^{d}}\; \big [ k_{y}^{2}-
k_{x}^{2} \big ] \;C(\vec{k},t)
\label{stress}
\end{equation}
\section{The scaling behavior}
Assuming simple scaling for the structure factor,
we write, for arbitrary space 
dimensionality $d$, 
\begin{equation}
\label{scaling}
C(\vec{k},t) = \prod_{i=1}^{d} R_i(t) F (\vec{X}, \tau(\gamma t))
\end{equation}
where the subscript $i$
labels the space directions with $i=1$ along the flow, 
$R_{i}$ is the average size of domains in the $i$-th direction, 
$\vec X$ is a vector of
components $X_i=k_i R_i(t)$ and $F$ is a scaling function 
\footnote{Our self-consistent approximation 
is 
equivalent to that made in the large-$N$ limit for vectorial systems 
\cite{Ma}. In this case 
simple scaling is not
verified for $\gamma=0$ when $\alpha=0$ \cite{Coniglio} and $\alpha \neq 0$
\cite{Bray} so that $C(\vec{k},t)$ has not the form (\ref{prima}) but 
can be written as $L^{d \xi(k/k_{m})}$, where $k_{m}$ is the position
of the maximum in the structure factor and $\xi$ is a function which
depends continuously on $k$
(multiscaling). 
In the case with shear the exact solution
is still lacking, so we cannot, in principle, rule out multiscaling. 
However, since simple scaling is the leading approximation in the regions
of the maxima of the structure factor, we can use it to obtain the correct
value of the growth exponents (apart from logarithmic corrections) because the
momentum integrals which define the observables are dominated by the maxima 
of $C(\vec{k},t)$.}. 
We also allow an explicit time dependence
of the structure factor through $\tau(\gamma t)$; notice that since
$C(\vec k,t)$ scales as the domains volume below the critical temperature,
$\tau $ must not introduce any further algebraic
time dependence in $C(\vec k,t)$. From the numerical results of the next
section, we will see that $F$ is a dumped periodic function of $\tau$.
Inserting the form (\ref{scaling}) of $C(\vec k,t)$
into Eq.~(\ref{eqM04}) we obtain :
\begin{eqnarray}
&&\gamma X_1F_2 = R_{1} R_{2}^{-1}\Bigg 
\{ \dot \tau \;\frac{\partial F}{\partial \tau}
+ \sum_{i=1}^{d} \bigg [ R_{i}^{-1} \dot R_{i} (F+X_{i} F_{i})\nonumber\\
&+&\! 2 \Big [ 1-S(t)\Big ]^{\alpha} \!\!R_{i}^{-2} X_{i}^{2}
\bigg ( \sum_{k=1}^{d} R_{k}^{-2}X_{k}^{2}-\!1 \!
+\!S(t) \bigg) F \bigg]\Bigg\}  
\label{ciccio}
\end{eqnarray}
where $F_i=\partial F/\partial X_i$ and a dot means
a time derivative.
Under the assumptions that $R_{1} \gg R_{i} \;\; (i=2,d)$ 
and $R_{i} \simeq \widetilde{R} \;\; (i=2,d)$ we can write
\begin{eqnarray}
\gamma X_1F_2 &=& \widetilde{R}^{-1} R_{1} \Bigg 
\{ \dot \tau \;\frac{\partial F}{\partial \tau}
+ R_{1}^{-1} \dot R_{1}(F+X_{1} F_{1}) \nonumber \\
&+&\widetilde{R}^{-1} \dot {\widetilde{R}} 
\sum_{i=2}^{d} 
\bigg [ (F+X_{i} F_{i})+
 2 \Big [ 1-S(t)\Big ]^{\alpha} \nonumber \\
&\times&\widetilde{R}^{-4}
X_{i}^{2}
\bigg ( \sum_{k=2}^{d} X_{k}^{2}-(1 -S(t))\widetilde{R}^{2} 
\bigg) F \bigg]\Bigg\}  
\label{ciccio2}
\end{eqnarray}
Since the l.h.s. of Eq.(\ref{ciccio2}) has no explicit algebraic time
dependence, 
one has the asymptotic solutions
\begin{eqnarray}
R_1(t)&\sim& \gamma \: t^{(5+2\alpha)/2(2+\alpha)}\nonumber \\
\widetilde{R}(t)&\sim& t^{1/2(2+\alpha)}\nonumber \\
\big(1-S(t)\big)&\sim& t^{1/(2+\alpha)}\\
\tau (\gamma t)&\sim& \log \gamma t \nonumber
\end{eqnarray}
The growth exponents in the flow and in the shear directions are
$z_{x}=(5+2\alpha)/2(2+\alpha)$ and $z_{y}=1/2(2+\alpha)$. 
We observe that  $\displaystyle z_{x}-z_{y}=\frac{4+2 \alpha}{2(2+\alpha)}=1$.
The value $z_{y}$ is the same
found in \cite{Bray}, in the leading approximation, in a model having a 
field-dependent mobility 
with vectorial conserved order parameter without shear. 
Increasing the values of $\alpha$, one obtains values smaller with respect 
to the case with constant mobility, when $z_{x}=5/4$
 and $z_{y}=1/4$.
The shear 
affects only the growth exponent $z_{x}$ 
which remains greater than $1$ for every real
and positive value of $\alpha$. 

The previous 
arguments can be used to establish the scaling properties of the
rheological coefficients. 
Inserting the form~(\ref{scaling}) into Eq.~(\ref{eqn5b})
we obtain 
\begin{eqnarray}
\Delta \eta (t)&\sim&
(\gamma t) ^{-(3+\alpha)/(2+\alpha)}
 \gamma ^{-(1+\alpha)/(2+\alpha)}\nonumber \\
&&\times \int X_1 X_2 F \left [ \vec X,\tau (t)\right ]d\vec X
\end{eqnarray}
Therefore, in the scaling
regime, for each value of $\gamma t$,   
the functions $\Delta \eta$ corresponding to
different values of $\gamma$ collapse each on the others if rescaled as
$\Delta \eta \!\rightarrow \!\gamma ^{(1+\alpha)/(2+\alpha)} \!\Delta \eta$. 
A similar 
analysis can be done for the normal stress. It is straightforward
to show that in the asymptotic regime
\begin{equation}
\Delta N_{1} \sim t^{-1/(2+\alpha)}
\int X_2^2 F \left [ \vec X,\tau (t)\right ] d\vec {X} 
\end{equation}
Setting $\alpha=0$, we recover the previous results for the case
with constant mobility\cite{Noi}.
\section{Results and discussion}
In this section we consider the numerical solution of Eq.~ (\ref{eqM03}).
We will present the results for the calculation of the average size
of domains and of the rheological indicators $\Delta \eta$ and
$\Delta N_{1}$, stressing the effects of a non-costant mobility.
\begin{figure*}
\resizebox{0.80\textwidth}{!}{%
  \includegraphics{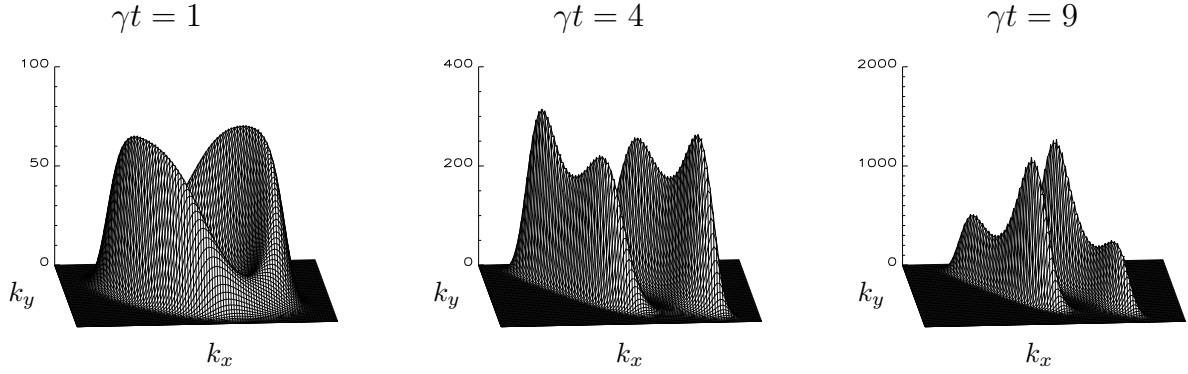}
}
\caption{The structure factor at consecutive times for $\gamma=0.01$
and $\alpha=1$. The $k_{x}$ coordinate assumes positive values on the right
of the picture, while the $k_{y}$ is positive towards the upper part of the
plane. The maximum value of $|k_{y}|$ is $0.8$. In the other direction,
$k_{x}$ varies in the range $[-0.8,0.8]$, $[-0.4,0.4]$ and 
$[-0.2,0.2]$, respectively. The positions of the peaks on the
left foil of the structure factor is $(-0.22,0.37)$ for the highest and
$(-0.08,-0.24)$ for the other at $\gamma t=4$. At $\gamma t=9$ the 
highest peak is located at $(-0.05,-0.23)$ and the other, at 
$(-0.10,0.36)$.}
\label{fig:1}     
\end{figure*}
We solved the equation of time evolution of the structure factor numerically
 in two dimensions, implementing a first-order Euler scheme with an
adaptive mesh. The starting configuration for the structure factor is a
constant value, which corresponds to a disordered state at very high
temperature. Eq.~  (\ref{eqM03})
has been solved for different values of $\gamma$.
In the following, results will be shown for the case $\gamma=0.01$. Similar
results have been obtained in the other cases. The two values 
$\alpha=1$ and $\alpha=2$ for the parameter $\alpha$ appearing in 
(\ref{newmob}) have been considered. 
At the
beginning the function $C(\vec{k},t)$ develops a circular volcano shaped
structure. This is then deformed, as consequence of shear, into 
an elliptic structure.
The sizes of the axes of the ellipse decrease in time at a 
different rate, this being larger in the $k_{x}$-direction.
During this evolution two dips start to develop in the volcano edge
until $C(\vec{k},t)$ is made of two foils. In Fig.~ 1 at $\gamma t=1$,
the shape of $C(\vec{k},t)$ representative of this stage of the
evolution is plotted. Later, on each foil, two 
well-formed peaks can be seen. At $\gamma t \simeq 4$
the peaks characterized by the larger values of $|k_{y}|$ prevail.
Observe that the structure factor is symmetric with respect to the change
$\vec{k} \rightarrow -\vec{k}$.  The peaks with the smaller $|k_{y}|$
corresponding to a more isotropic configuration of 
domains, grow faster than the others until they 
prevail as it can be seen in Fig.~ 1 at  $\gamma t=9$.
These peaks continue to prevail along all
the time evolution. What we observed is that the peaks continue always to grow 
and that the difference of their heights as a function of the shear strain
$\gamma t$ is an increasing function being modulated by 
dumped oscillations. 

In order to have information about the growth of domains, we computed the 
typical domain size as 
\begin{equation}
R_{x}(t)=\left(\frac{\int d \vec{k} C(\vec{k},t)}
{\int d \vec{k} k_{x}^{2} C(\vec{k},t)} \right)^{1/2}
\end{equation}
and the same for the other direction. The values of $R_{x}$ and $R_{y}$ are 
plotted in Fig.~ 2 as function of the shear strain $\gamma t$ for 
$\alpha=1$ and $\alpha=2$. We plotted also the values for the case with
constant mobility ($\alpha=0$). 
Some comments are in order here. The
asymptotic behavior is the one expected through the previous scaling
analysis: for $\alpha=1$ one has $R_{x} \sim t^{7/6}$ and $R_{y} \sim t^{1/6}$;
for $\alpha=2$, $R_{x} \sim t^{9/8}$ and $R_{y} \sim t^{1/8}$.
 The growth exponents in both the directions are decreasing
functions of the mobility exponent $\alpha$. This is reasonable since the
mobility (\ref{newmob}) becomes smaller and smaller when $\alpha$
increases, if $(1-\phi^{2})<1$.
Another
consideration is about the superimposed log-time periodic oscillations. In the
case with constant mobility  these oscillations have an apparently
constant amplitude.
For non-zero values of $\alpha$ these oscillations are dumped. We will see that this feature is common to all the observables and will be discussed
later.
\begin{figure}
\resizebox{0.49\textwidth}{!}{%
  \includegraphics{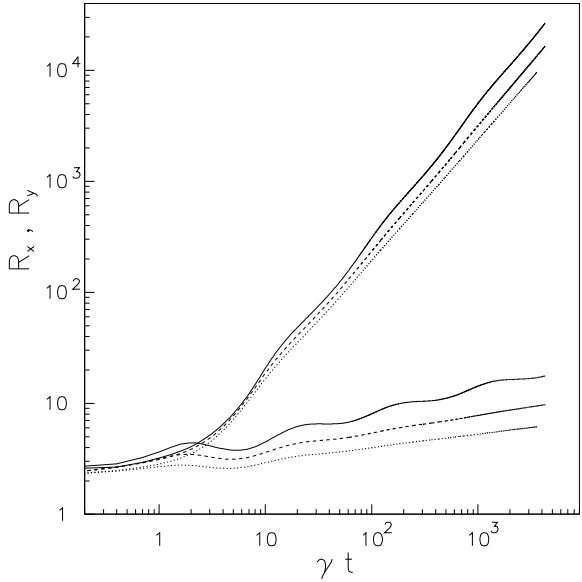}
}
\caption{The typical size of domains as function of the strain 
$\gamma t$ in the flow
and in the shear directions for different values of $\alpha$:
$\alpha=0$ (full line), $\alpha=1$ (dashed line),  
$\alpha=2$ (dotted line).}
\label{fig:2}     
\end{figure}

We turn now to the study of the rheological behavior of the 
system. The external
velocity field causes additional stresses on the mixture. Important indicators
are the excess viscosity and the first normal stress. We calculated 
numerically $\Delta \eta$ and $\Delta N_{1}$ using their definitions
(\ref{eqn5b}) and (\ref{stress}), respectively. The results are shown in
Fig.~ 3 and Fig.~ 4. 
\begin{figure}
\resizebox{0.49\textwidth}{!}{%
  \includegraphics{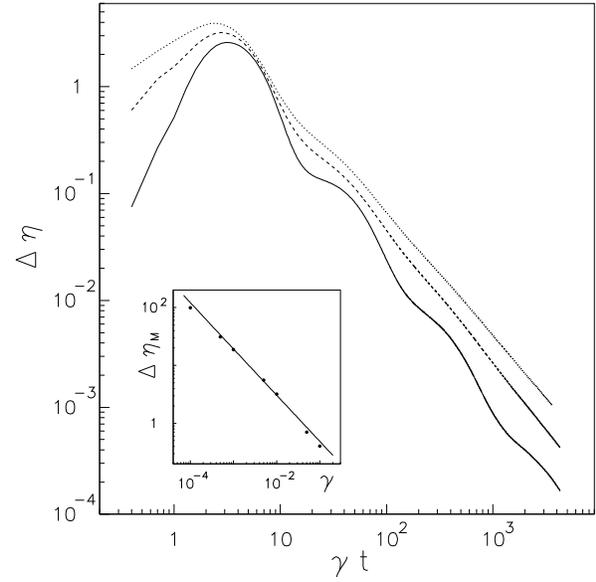}
}
\caption{Plots of the excess viscosity vs the strain $\gamma t$
for different values of $\alpha$:
$\alpha=0$ (full line), $\alpha=1$
(dashed line), $\alpha=2$ (dotted line). 
The inset shows the maxima of $\Delta \eta$
as function of $\gamma$ for the case $\alpha=1$.
The slope of the straight line is 0.8.}
\label{fig:3}     
\end{figure}
\begin{figure}
\resizebox{0.49\textwidth}{!}{%
  \includegraphics{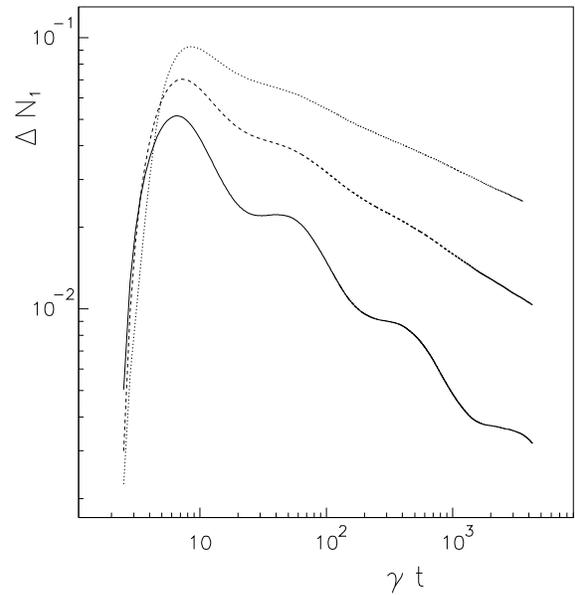}
}
\caption{The first normal stress  as function of the strain 
$\gamma t$ for different values of $\alpha$:
$\alpha=0$ (full line), $\alpha=1$
(dashed line), $\alpha=2$ (dotted line).}
\label{fig:4}     
\end{figure}
The excess viscosity reaches its maximum at the onset
of the scaling when the domains are expected to be maximally
stretched in the flow direction and the 
structure of $C(\vec{k},t)$ is the one shown at $\gamma t=4$ in Fig.~ 1.
According to our analysis of the scaling behavior, we expect 
the excess viscosity to scale with $\gamma$
as $\Delta \eta \sim \gamma^{-(1+\alpha)/(2+\alpha)}$ for fixed value of
$\gamma t$. In the inset of Fig.~ 3 we report for the case $\alpha=1$
the dependance of $\Delta \eta_{M}$ on $\gamma$. We find an exponent $0.8$
slightly larger than the expected $2/3$. The reason can be related to the fact 
that the excess viscosity reaches the maximum before the asymptotic regime
is fully realized.
Then $\Delta \eta$ decreases as a consequence of the
dissipation of the elastic energy stored 
by domains which start to burst when they
are stretched furtherly. Therefore more isotropic patterns form and the 
typical structure of the function $C(\vec{k},t)$ is the one at 
$\gamma t=9$ in Fig.~ 1. The excess viscosity decreases with a power law
behavior which is consistent with the predicted exponent 
$-(3+\alpha)/(2+\alpha)$. The first normal stress reported in Fig.~ 4
decreases in
time according to the exponent $-1/(2+\alpha)$ after reaching a maximum.
The amplitudes of all quantites plotted in Figures 2, 3, 4 are 
modulated by dumped log-time
oscillations \cite{Sornette}.
We believe that the physical explanation for the dumping of
oscillations 
may be found in the vanishing value of the mobility at equilibrium.
In the case with constant mobility the origin of the
oscillations is related to a cyclical mechanism of elongation and bursting of 
domains, which allows to store and dissipate elastic energy in the 
system \cite{Noi}. In the present case,
during the time evolution, the decreasing values of $\Gamma (\phi)$ suppress
diffusion in the bulk phase and inhibit the growth of small bubbles coming
from bursting. Therefore, they cannot be stretched too much by the flow. 
In this way it is more difficult to store elastic energy in the system
and the excess viscosity
can increase only of a small amount.
This mechanism of growth inihibition
becomes stronger and stronger in the course of evolution causing the observed
dumping of oscillations. 

In conclusion, we have studied the phase separation of a binary mixture with
field-dependent mobility in shear flow. We proved that dynamical scaling
holds for this system. There are different growth exponents in the flow and
in the shear directions which depend on the mobility exponent $\alpha$.
The difference in growth exponents is always $1$.
All the physical observables have amplitudes decorated by dumped oscillations
which are periodic in logarithmic time. We made a guess about the origin of this behavior. It would be an important endeavour to study this system in
direct simulation of Eq.~ (\ref{eqM01}) to deeply understand this phenomenon.
\begin{acknowledgement}
We thank Federico Corberi for valuable discussions about the subject
of this work.
\end{acknowledgement}

\end{document}